\documentclass[twocolumn,showpacs,amsmath]{revtex4}
\usepackage{graphicx}%
\usepackage{dcolumn}
\usepackage{amsmath}
\usepackage{here}
\begin{document}

\title{Partially incoherent optical vortices in self-focusing nonlinear media}

\author{Chien-Chung Jeng$^1$, Ming-Feng Shih$^1$, Kristian Motzek$^{2,3}$, and Yuri Kivshar$^3$}

\affiliation{$^1$Physics Department, National Taiwan University,
Taipei, 106, Taiwan \\ $^2$ Institute of Applied Physics,
Darmstadt University of Technology, D-64289 Darmstadt, Germany\\
$^3$ Nonlinear Physics Group, Research School of Physical Sciences
and Engineering, Australian National University, Canberra ACT
0200, Australia}

\begin{abstract}
We observe stable propagation of spatially localized single- and
double-charge optical vortices in a self-focusing nonlinear
medium. The vortices are created by self-trapping of partially
incoherent light carrying a phase dislocation, and they are
stabilized when the spatial incoherence of light exceeds a certain
threshold. We confirm the vortex stabilization effect by numerical
simulations and also show that the similar mechanism of
stabilization applies to higher-order vortices.
\end{abstract}

\maketitle

Vortices are fundamental localized objects which appear in many
branches of physics, ranging from liquid crystals to superfluids
and Bose-Einstein condensates~\cite{pismen}. In optics, vortices
are associated with phase dislocations (or phase singularities)
carried by diffracting optical beams~\cite{soskin}. When such
vortices propagate in {\em self-defocusing} nonlinear media, the
vortex core with a phase dislocation becomes self-trapped, and the
resulting {\em stationary singular beam} is known as {\em an
optical vortex soliton}~\cite{book}. Such vortex solitons have
been generated experimentally (e.g. by using a phase mask) within
broad diffracting beams in different types of defocusing nonlinear
media~\cite{swartz,exp1,exp2,exp3,review}. Optical vortex solitons
demonstrate many common properties with the vortices observed in
superfluids and, more recently, in Bose-Einstein
condensates~\cite{bec}.

In contrast, optical vortices become highly unstable in
self-focusing nonlinear media. Indeed, when a ring-like optical
beam with zero intensity at the center and a phase
singularity~\cite{kruglov} propagates in a self-focusing nonlinear
medium, it always decays into several fundamental solitons flying
off the main ring~\cite{firth}. This effect has been observed
experimentally in different nonlinear systems, including the
saturable Kerr-like nonlinear media~\cite{tikh}, biased
photorefractive crystals~\cite{exp3}, and quadratic nonlinear
media~\cite{chi2} in the self-focusing regime. This effect has
also been predicted to occur in many other physical systems
including attractive Bose-Einstein condensates~\cite{saito}.

A number of recent theoretical papers~\cite{cubic}, including the
rigorous studies of the vortex stability~\cite{buryak}, suggest
that the stable propagation of spatial and spatiotemporal
vortex-like stationary structures may become possible in the
models with competing nonlinearities in the presence of a large
higher-order defocusing nonlinearity, but {\em no realistic
physical systems} to support these theoretical findings have been
found so far. Thus, the main question remains open: {\em Can
stable optical vortices readily be observed in self-focusing
nonlinear media?}

The main purpose of this Letter is to demonstrate, for the first
time to our knowledge, that {\em stable propagation of spatially
localized optical vortices} in a self-focusing photorefractive
crystal can indeed be observed provided such vortices are created
by partially incoherent light carrying a phase dislocation. In
particular, we show, both experimentally and theoretically, that
single- and double-charge optical vortices can be stabilized in
self-focusing nonlinear media when the value of the spatial
incoherence of light exceeds a certain threshold, and these
vortices are readily observed in experiment as stationary
self-trapped structures propagating for many diffraction lengths.

We should mention that the generation and properties of singular
optical beams created by partially incoherent light is an
important issue which is a subject of a current active research
even in linear optics (see, e.g, Ref.~\cite{gbur} and references
therein). Our results demonstrate that partially incoherent
singular beams can exist in self-focusing  nonlinear media being
stabilized by the light incoherence effect.


First, we present our experimental results. The main purpose of
our experiments conducted for a biased photorefractive medium is
to generate partially incoherent vortices and vortex solitons, and
then inspect their stability in such a self-focusing nonlinear
medium. Being driven by the earlier results of Anastassiou {\em et
al.}~\cite{inst_exp} who demonstrated the suppression of
modulational instability for the stripe spatial solitons created
by partially incoherent light, we try to understand whether the
other type of the nonlinearity-driven instability, the so-called
{\em azimuthal instability}, can be suppressed by reducing the
coherent properties of optical vortex beams.

\begin{figure}
\centerline{\includegraphics[width=3.3in]{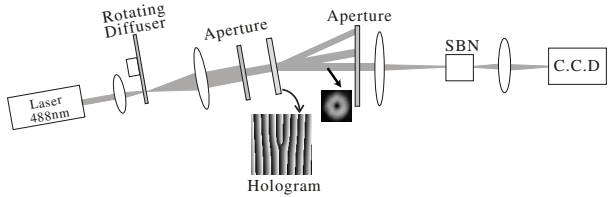}}
\caption{Schematic of the experimental setup for the observation
of the partially incoherent optical vortices; SBN: Strontium
Barium Niobate crystal.} \label{fig_exp1}
\end{figure}

The experiment (the setup is shown in Fig.~\ref{fig_exp1}) is
conducted with a biased photorefractive SBN crystal (a x b x c = 5
mm x 10mm x 5mm). First, a cw laser light beam (at 488 nm) of the
extraordinary polarization is made partially incoherent by passing
it through a lens and then through a rotating diffuser. The
rotating diffuser introduces {\em random-varying phase and
amplitude} on the light beam every 1 $\mu$s, which is much shorter
than the response time (about 1 s) of the crystal under our
experimental light illumination. By adjusting the position of the
diffuser to near (away from) the focal point of the lens in front
the diffuser, we can increase (decrease) the degree of the light
coherence. We collect the light after the rotating diffuser by a
second lens and then pass is through a computer-generated hologram
to imprint a vortex phase (with a single-, for $m=1$, or double-,
for $m=2$, charge) on the light beam. Since the partially
incoherent light beam can be considered as {\em a superposition of
many mutually-incoherent light beams}, the first-order diffracted
light beam after the hologram becomes a superposition of many
mutually-incoherent vortex beams. We then focus and launch the
partially coherent vortex beam into the SBN crystal along its
a-axis. The crystal is illuminated by a halogen lamp from its top
side as necessary for photorefractive screening solitons. The
total power of the vortex beam is of 0.17$\mu$W, which results in
the nonlinearity of the photorefractive crystal falling into the
Kerr region for the peak intensity of the vortex beam to the
background intensity is much less than unity. Then, a lens is used
to project the images at the input and output faces onto a CCD
camera.

\begin{figure}
\centerline{\includegraphics[width=3.5in]{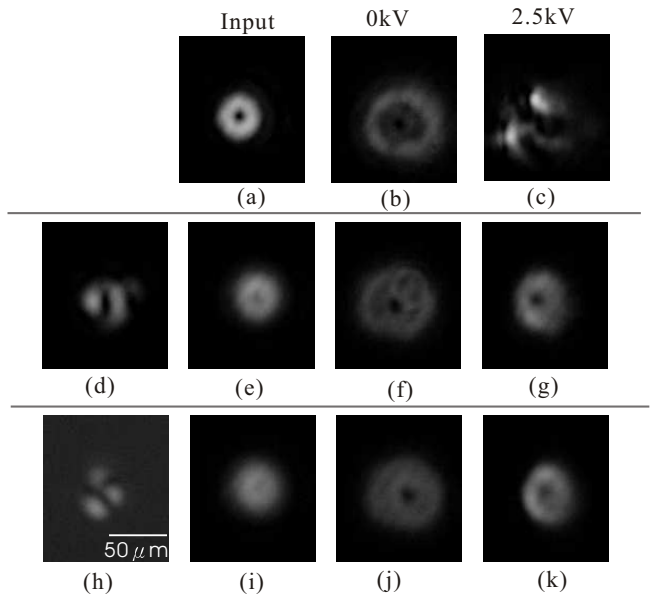}}
\caption{Experimental results for the intensity distribution of a
single-charge vortex beam ($m=1$) for different degrees of
coherence (top: coherent, middle: less coherent, bottom: least
coherent). At input face, (a) is coherent light, (e) and (i) are
partially incoherent, but (i) is more incoherent than (e) by
comparing the speckle pattern shown in (d) and (h). At output
face, (b) (f) and (j) show the diffraction when the nonlinearity
is off, and (c), (g) and (k) shows the incoherence stabilize the
vortex soliton when voltage of 2.5 kV is applied.}
\label{fig_exp2}
\end{figure}

\begin{figure}
\centerline{\includegraphics[width=3.5in]{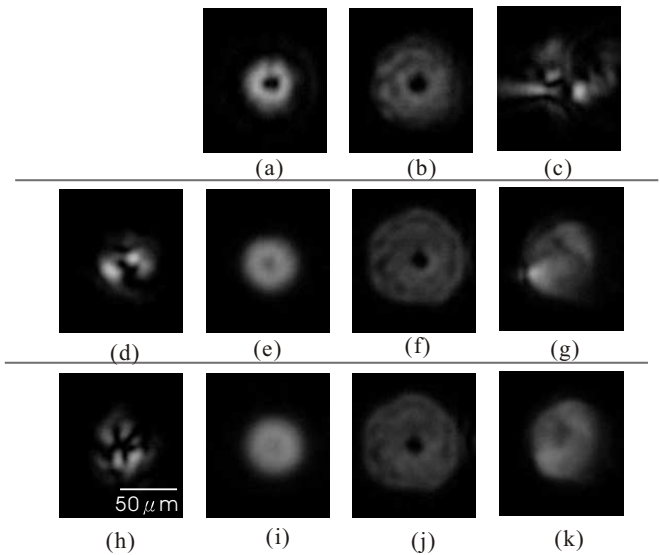}}
\caption{Experimental results for a double-charge vortex beam
($m=2$) for different degrees of coherence (top: coherent, middle:
less coherent, bottom: least coherent). At input face, (a) is
coherent light, (e) and (i) are partially incoherent, but (i) is
more incoherent than (e) by comparing the speckle pattern shown in
(d) and (h). At output face, (b) (f) and (j) show the diffraction
when the nonlinearity is off, and (c), (g) and (k) shows the
incoherence stabilize the vortex soliton when voltage of 3 kV is
applied. } \label{fig_exp3}
\end{figure}

Before showing that reducing the degree of coherence can stabilize
the vortex beam propagating in a self-focusing medium, we
reproduce the experiment that the coherent single-charge($m=1$)
vortex light beam cannot stably propagate in a self-focusing
medium~\cite{tikh,exp3}. We first remove the diffuser from the
experimental setup. The vortex beam at the input face of the
crystal is shown as Fig.~\ref{fig_exp2}(a). With zero biasing
voltage, Fig.~\ref{fig_exp2}(b) shows the natural diffraction of
the vortex light beam. While a 2.5 kV biasing voltage is applied
on the photorefractive crystal creating a Kerr-type self-focusing
nonlinear medium, the vortex beam breaks up into two pieces
[Fig.~\ref{fig_exp2}(c)]. This vortex break-up observed in a
self-focusing medium is due to the azimuthal instability, and it
has been theoretically and experimentally demonstrated
previously~\cite{firth,tikh}. We then put back the rotating
diffuser and adjust it to a suitable position. The degree of
coherence of the vortex light beam can be estimated by the speckle
size at the input face of the crystal [Fig.~\ref{fig_exp2}(d)]
when we stop the diffuser from rotating. As a voltage of 2.5 kV is
applied on the crystal, Fig.~\ref{fig_exp2}(g) clearly shows that
the vortex light beam is stabilized by the reduction of the degree
of coherence though two very unclear bright spots still can be
seen on the opposite sides (top and bottom) of the ring-like
intensity distribution. As we move the rotating diffuser further
away from the focal point of the lens to make the light more
incoherent [indicated by Fig.~\ref{fig_exp2}(h)] and apply a
voltage of 2.5 kV on the crystal, Fig.~\ref{fig_exp2}(k) shows the
generated stable {\em partially incoherent vortex soliton} at the
output face of the crystal.

We notice that the vortex stabilization by the light incoherence
can possibly be explained by employing simple physics. Indeed,
after some propagation of the partially incoherent beam with an
imprinted phase, we observe that the intensity at the center of
the beam does not vanish. This means that the incoherent vortex
can be decomposed into many mutually incoherent vortex beams with
not only their phases randomly distributed but also their central
positions offset from each others. In this way, the core of a
composite vortex beam will be filled by some light, and the index
change at the beam center will become nonzero. We believe that
this effect contributes strongly to the vortex stabilization.

We continue with the experiments for the double-charge vortex
beams generated by a phase mask with the $m=2$ dislocation.
Figures~\ref{fig_exp3}(a-c) show that a coherent double-charge
vortex light beam cannot stably propagate in a self-focusing
medium (we applied 3 kV here) as well, and it breaks up into
pieces as been observed in a self-focusing atomic
vapor~\cite{tikh}. When we make the light more incoherent, the
double-charge vortex becomes more stable, as shown in
Figs.~\ref{fig_exp3}(g) and (k). However, we could not make a
double-charge vortex soliton here. The double-charge vortex beam
diffracts more than a single-charge vortex beam and, therefore, it
requires higher nonlinearity (or higher voltage) to form the
vortex soliton, but our crystal begins to arc at 3.5 kV.
Nevertheless, these two experiments show that the reduction of the
degree of coherence of the light beam can indeed stabilize the
single- or double-charge (or even higher-charge) vortex solitons
propagating in a self-focusing medium.


\begin{figure}
\centerline{\includegraphics[width=3.3in]{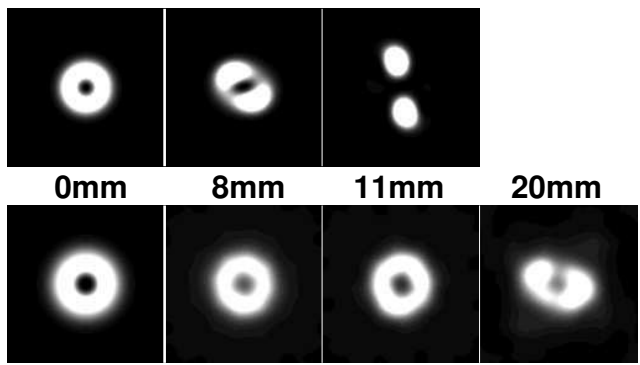}}
\caption{Numerical results showing the comparison between the
evolution of the vortex created by a coherent (upper row) and
partially incoherent (lower row) light ($\theta_0=0.38$). The
vortex stabilization by partial incoherence is clearly seen.}
\label{fig_theory2}
\end{figure}

\begin{figure}
\centerline{\includegraphics[width=3.3in]{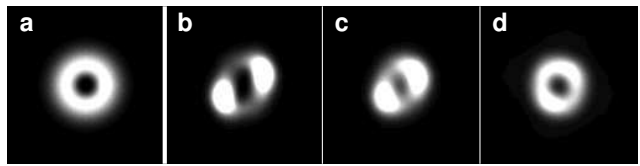}}
\caption{Numerical results showing the stabilization of the vortex
with growing incoherence: (a) input intensity, (b) vortex after
9mm of propagation for the coherent case, (c) vortex after 9mm for
the partially incoherent case, $\theta_0=0.14$, and (d) vortex
after 9mm for the partially incoherent case, $\theta_0=0.29$.}
\label{fig_theory}
\end{figure}

We have also studied numerically the propagation of partially
incoherent optical vortices in a photorefractive medium, employing
the coherent density approach~\cite{IncNum,inst_exp}. The coherent
density approach is based on the fact that partially incoherent
light can be described by a superposition of mutually incoherent
light beams that are tilted with respect to the $z$-axis at
different angles. One thus makes the ansatz that the partially
incoherent light consists of many coherent, but mutually
incoherent light beams $E_j$: $I=\sum_j |E_j|^2$. By setting
$|E_j|^2=G(j\vartheta) I$, where
\begin{equation}
G(\theta)=\frac{1}{\sqrt{\pi}\theta_0} e^{-\theta^2/\theta_0^2}
\label{eq}
\end{equation}
 is the angular power spectrum, one obtains a partially
incoherent light beam whose coherence is determined by the
parameter $\theta_0$, i.e. less coherence means bigger $\theta_0$.
Here, $j\vartheta$ is the angle at which the $j$-th beam is tilted
with respect to the $z$-axis. For our simulations we used a set of
1681 mutually incoherent vortices, all initially tilted at
different angles. To simulate the photorefractive nonlinearity we
use a simple model which predicts that the refractive index change
is approximately given by $I/(1+I)$.

Figure~\ref{fig_theory2} shows a comparison between the
propagation of the vortices generated by coherent and partially
incoherent light (for $\theta_0=0.38$). Increasing the incoherence
further leads to the case where the vortex beam radiates off a lot
of its intensity before the azimuthal instability can set in.
Larger values of incoherence (i.e. larger values of $\theta_0$)
correspond to a complete suppression of the vortex instability,
and this confirms the experimental results presented above.

Figures~\ref{fig_theory}(a-d)  show our numerical results for the
propagation of an input Gaussian beam carrying a phase dislocation
[(a)] after the total propagation (9 mm) in a nonlinear medium for
the coherent light [(b)] and two different partially incoherent
beams [(c,d)], corresponding to the values $\theta_0=0.14$ and
$\theta_0=0.29$, respectively. The most obvious difference to the
scenario of the propagation the a coherent vortex is that the
vortex decay undergoes a visible delay when the degree of
incoherence grows. Furthermore, in the incoherent case the vortex
changes its profile only very slowly as it propagates and thus can
be considered as being in a transition stage between the decay and
stabilization.

\begin{figure}
\centerline{\includegraphics[width=3.3in]{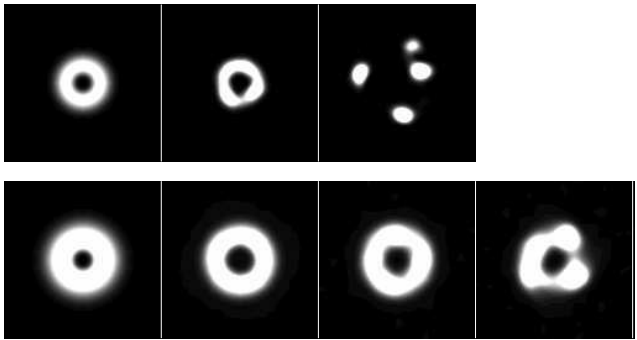}}
\caption{Propagation of a double-charge vortex in (top row) the
coherent case at $z=$0, 8.6, and 14.5 mm, and (bottom row) in the
partially incoherent case with $\theta_0=0.35$ at $z=$0, 8.6,
14.7, and 18.8 mm.} \label{fig_theory3}
\end{figure}

Finally, we have studied numerically the propagation of a
double-charge vortex beam in both coherent and partially
incoherent cases, as shown in Fig.~\ref{fig_theory3}. The initial
perturbation in both cases have been chosen to be very small in
order to obtain a clearer picture of the instability, therefore
the propagation distances are quite long. In numerics, since we
have a possibility to observe long propagation distances, we did
not find that the vortices can be completely stabilized by
increasing incoherence. When using the values of $\theta_0$ far
above the value $\theta_0=0.35$ used in Fig.~\ref{fig_theory3}, we
observe that the vortex just radiates off a lot of its intensity
and then decays. The problem is obviously that the vortices are
not only incoherent along the angular, but also the radial
direction. We believe that the experiment would confirm this
observation if a longer crystal was available.


In conclusion,  for the first time to our knowledge, we have
observed the stable propagation of single- and double-charge
optical vortices in a self-focusing nonlinear medium. The vortices
have been created by partially incoherent light beams carrying a
phase dislocation and propagating in a photorefractive nonlinear
medium. The vortex azimuthal instability in a self-focusing
nonlinear medium was suppressed for the light incoherence above a
critical value. The experimental results have been confirmed by
numerical simulations which also provide an insight into the
physical mechanisms of the vortex stabilization.

Yuri Kivshar thanks the Physics Department of the Taiwan
University for hospitality. This work was supported by a
collaboration program between the Australian Academy of Science
and the National Science Council of Taiwan, and the Australian
Research Council.

\end{document}